\documentclass[onecolumn]{aastex6}
\def\NEW#1{{\textcolor{black}{#1}}}
\def\NEWER#1{{\textcolor{black}{#1}}}

\renewcommand{\vec}[1]{\mathbf{#1}}

\def\hscale{1}

\usepackage{hyperref}
\usepackage[all]{hypcap}

\begin{document}
\title{Small-scale dynamo simulations: Magnetic field amplification in exploding granules and the role of deep and shallow recirculation}
\shorttitle{Small-scale dynamo simulations}

\author{M. Rempel\altaffilmark{1}}

\shortauthors{Rempel}

\altaffiltext{1}{High Altitude Observatory,
    NCAR, P.O. Box 3000, Boulder, Colorado 80307, USA}

\email{rempel@ucar.edu}

\begin{abstract}
We analyze recent high resolution photospheric  small-scale dynamo simulations that were computed with the MURaM radiative MHD code.
We focus the analysis on newly forming downflow lanes in exploding granules since they show how weakly magnetized regions in
the photosphere (center of granules) evolve into strongly magnetized regions (downflow lanes). We find that newly formed downflow
lanes exhibit initially mostly a laminar converging flow that amplifies the vertical magnetic field embedded in the granule from a few 10~G 
to field strengths exceeding 800~G. This results in extended magnetic sheets that have a length comparable to granular 
scales. Field amplification by turbulent shear happens first a few 100~km beneath the visible layers of the photosphere. Shallow recirculation
transports the resulting turbulent field into the photosphere within minutes, after which the newly formed downflow lane shows a mix of 
strong magnetic sheets and turbulent field components. We stress in particular the role of shallow and deep recirculation for the organization and 
strength of magnetic field in the photosphere \NEW{and discuss the photospheric and sub-photospheric energy conversion associated with the small-scale dynamo process.}
\NEWER{While the energy conversion through the Lorentz force depends only weakly on the saturation field strength (and therefore deep or shallow recirculation), 
it is strongly dependent on the magnetic Prandtl number.}
We discuss the potential of these findings for further constraining small-scale dynamo models through high resolution observations.
\end{abstract}


\received{}
\accepted{}

\maketitle
\section{Introduction}
Observations suggest that small-scale magnetic field in the solar photosphere is mostly independent from the strength
of nearby network field \citep{Ishikawa:2009:cmp_QS_plage,Lites:2011:hinode_ssd,Lamb:etal:2014} as well as independent of the solar cycle 
\citep{Buehler:2013:cyc_dep,Lites:2014:cycle_dep}. This supports the view that the origin of small-scale magnetism is due to a 
small-scale dynamo that operates independently from the large-scale dynamo responsible for the solar cycle. A small-scale dynamo
as origin of the quiet sun magnetic field was first suggested by \citet{Petrovay:1993:SSD} and numerical simulations in incompressible 
setups conducted by \citet{Cattaneo:1999} further supported that concept. \citet{Stein:2003:review} pointed out that there is only
very little local re-circulation in the photosphere and they suggested that even the small-scale dynamo has to operate throughout the convection zone on a global scale.
\citet{Voegler:Schuessler:2007} presented the first realistic solar MHD simulations with radiative transfer and demonstrated that
an excited small-scale dynamo can exist in the photosphere in spite of low local recirculation (they used a bottom boundary condition that
did not allow for a Poynting flux in upflows, which essentially isolates the photosphere from the rest of the convection zone). 
However, the saturation field strength ($\langle\vert B_z\vert\rangle(\tau=1) \approx 25$~G) fell short of 
observations by about a factor of $2-3$ \citep{Danilovic:2010:zeeman_dynamo}. \citet{Rempel:2014:SSD} presented models
with reduced numerical diffusivities and different bottom boundary conditions that emulate the presence of a deep magnetized
convection zone. It was found that just reducing numerical diffusivities alone does not increase the saturation field strength sufficiently
beyond the values found by \citet{Voegler:Schuessler:2007}, instead the bottom boundary condition plays a key-role.
A saturation field strength consistent with observational constraints \citep{Rempel:2014:SSD,Danilovic:2016:SSD_Power} requires 
a setup that has either complete recirculation within the simulation domain (closed bottom boundary), or a bottom boundary that 
emulates the presence of a deep magnetized convection zone through the presence of horizontal magnetic field in upflows 
as previously conjectured by \citet{Stein:2003:review} . Solutions that are consistent with observations 
($\langle\vert B_z\vert\rangle(\tau=1) \approx 60-80$~G) require a subsurface
RMS field strength that is about  $0.5-1\,B_{\rm eq}$, with the equipartition field strength $B_{\rm eq}=\sqrt{4\pi\varrho}\,V_{\rm RMS}$, 
\citep{Rempel:2014:SSD}.  Such subsurface field strength values are consistent
with those found in deep seated small-scale dynamo simulations that reach to the base of the solar convection zone and therefore
account for full recirculation, but do not include the upper layers of the convection zone \citep{Hotta:2015:SSD}.

Overall the above research suggests that the efficiency of recirculation plays a central role in determining the photospheric saturation
values of the small-scale dynamo, i.e. quiet sun magnetism is not due to a  ``local" dynamo operating in the photosphere, it is more a reflection of a deep
seated small-scale dynamo where the deeper convection zone plays a key role in shaping the photospheric appearance of small-scale magnetism.

Since both resolution (which in combination with either explicit or implicit magnetic diffusivity determines the super-criticality of the dynamo during its kinematic phase) 
and boundary conditions influence the saturation field strength of a dynamo simulation, it is not surprising that different models do not necessarily show comparable 
saturation levels. For example \citet[][see, Figure 16 therein]{Kitiashvili:2015:SSD} found photospheric 
values of $\langle\vert B_z\vert\rangle$ and $\langle B_h\rangle$ in the $10-20$~G range, which is lower than \citet{Voegler:Schuessler:2007}; 
\citet{Khomenko:2017:Biermann} found a photospheric saturation field strength of $\langle \vert \vec{B}\vert\rangle\approx 100$~G, which is about half way 
between the results of \citet{Voegler:Schuessler:2007} and \citet{Rempel:2014:SSD}. The role of resolution and boundary conditions was not further investigated
in \citet{Kitiashvili:2015:SSD} and \citet{Khomenko:2017:Biermann}, which makes a cross-comparison of all these models challenging.

\NEW{
In addition to the role of recirculation, the role of the magnetic Prandtl number ($P_{\rm m}$) on the dynamo efficiency has been studied in great detail. The latter is
of particular interest for the solar dynamo since $P_{\rm m}$ reaches values as low as $10^{-5}$ in the photosphere. \citet{Schekochihin:etal:2005,Schekochihin:etal:2007}
considered simulations of forced turbulence to study the onset of dynamo action in the kinematic phase. \citet{Schekochihin:etal:2007} and \citet{Iskakov:2007:SSD_low_Pm} concluded that the fluctuation dynamo 
does exist also in the low $P_{\rm m}$ regime, but with a critical magnetic Reynolds number about a factor 3 larger compared to the high $P_{\rm m}$ regime. \NEWER{While these direct numerical simulations could not
achieve $P_{\rm m}$ values lower than $0.1$ to $0.01$, the asymptotic limit of very low  $P_{\rm m}$  has been studied in a more idealized model by  \citet{Bouldurev:Cattaneo:2004}.
They found that small-scale dynamo action remains possible for rough velocity fields (low $P_{\rm m}$), while the
critical magnetic Reynolds number is larger by a factor of a few (see also the review of \citet{Tobias:Cattaneao:Boldyrev:SSD_review} for further
discussion).} Studies of the role of $P_{\rm m}$ for the saturated small-scale dynamo were conducted by
\citet{Brandenburg:2011:SSD_low_Pm,Brandenburg:2014:Pm}. They found that the dynamo efficiency (ratio of work against Lorentz force and the available driving through forcing terms) 
\NEWER{depends upon the magnetic Prandtl number}, suggesting that $P_{\rm m}$ strongly influences the ratio of resistive to viscous energy dissipation, with resistive dissipation being dominant
in the low $P_{\rm m}$ regime.}

\NEW{
$P_{\rm m}$ dependence studies are much more challenging for convective dynamo setups representing conditions in the solar photosphere. \citet{Bushby:Favier:2014} considered
convective small-scale dynamos on scales of granulation and meso-granulation. They used explicit viscosity and magnetic resistivity and found excited dynamos only for $P_{\rm m}$ 
values larger than $0.5$. Many of the recent solar-like simulations rely at least partially on numerical diffusivities using either implicit or explicit subgrid-scale models, which makes a 
quantification of $P_{\rm m}$ challenging. Since numerical diffusivities can vary substantially based on the local structure of velocity and magnetic field, $P_{\rm m}$, if defined locally,
is highly intermittent and varies substantially throughout the simulation domain. \citet{Pietarila-Graham:etal:2010:SSD} analyzed dynamo simulations similar to those presented by 
\citet{Voegler:Schuessler:2007} and defined a global numerical $P_{\rm m}$ based on the kinetic and magnetic energy Taylor microscale and found values in the $0.8-2$ range. \citet{Thaler:Spruit:2015:SSD_Pm}
found an excited small-scale dynamo for $P_{\rm m}=5$ in their setup (about neutral growth for $P_{\rm m}=2$), where $P_{\rm m}$ was defined as a pre-factor that scales magnetic hyperdiffusivity relative to hyperviscosity (which can be also
considered a global definition of $P_{\rm m}$). In contrast to that \citet{Kitiashvili:2015:SSD} and  \citet{Khomenko:2017:Biermann} considered local definitions of $P_{\rm m}$ and found that those values vary
substantially by several orders of magnitude. Most of the induction as well as magnetic energy density was however associated with local $P_{\rm m}$ slightly smaller than $1$.
}

\NEW{ In the first part of this paper} we present a study that aims at identifying observables that could help to further constrain the role of recirculation. To this end we analyze
and compare 2 simulations from \citet{Rempel:2014:SSD} that have the same numerical resolution, but differ in their treatment of the bottom boundary
condition. In particular we focus
on newly forming downflow lanes in exploding granules that show how the most weakly magnetized regions in the photosphere (center of granules) 
evolve into the most strongly magnetized regions (downflow lanes). That process is in particular sensitive to the ``seed field" that is present in the center
of granules and brought into the visible layers of the photosphere through recirculation. While the outer regions of granules show mostly
turbulent field that is transported back into the photosphere due to local recirculation (upflow/downflow mixing), the centers of granules are expected
to exhibit field that is brought up from deeper layers of the photosphere.
We study primarily exploding granules since they provide the most undisturbed view on the field amplification process, the described picture 
is however representative for the photosphere as a whole. 
\NEW{
In the second part of paper we focus on the photospheric and sub-photospheric energy conversion and discuss how the dynamo efficiency depends
on shallow/deep recirculation as well as the magnetic Prandtl number. In particular we present here an extension of the work by \citet{Rempel:2014:SSD}
by considering also models that use "numerical" $P_{\rm m}$ with values far from unity. We focus here in particular on the findings of 
\citet{Brandenburg:2011:SSD_low_Pm,Brandenburg:2014:Pm} on the role of $P_{\rm m}$ for the energy dissipation in a saturated dynamo.
}

\section{Simulation data}
We analyze a photospheric small-scale dynamo simulation that was already presented in \citet{Rempel:2014:SSD}.
We focus on the model `O4a'  with 4~km grid spacing in a 6.144~Mm wide and 3.072 Mm deep domain. The 
photosphere is located about 700~km beneath the top boundary. The setup reaches at an optical depth of unity a 
mean vertical field strength of about $\langle \vert B_z\vert\rangle\sim 85$~G, which corresponds to 
$\langle \vert \vec{B}\vert\rangle\sim165$~G, and  $B_{\rm RMS}\sim 250$~G. In the following we refer to this simulation as
${\rm SSD}_{\rm Deep}$. The saturation field strength of $85$~G is moderately
stronger than the value of $60$~G inferred from Hinode observations \citet{Danilovic:2016:SSD_Power} through forward 
modeling of spectro-polarimetric signatures. Recently del Pino Aleman et al. (2018, in prep.) found that the magnetization of a \citet{Rempel:2014:SSD} model 
with a surface mean field strength of about $160$~G is consistent with Hanle-depolarization observed in the Sr I 460.7 nm line.
The model ${\rm SSD}_{\rm Deep}$ uses an open bottom boundary
condition that mimics a deep magnetized convection zone by allowing for the presence of horizontal magnetic field
in upflow regions (deep recirculation). The boundary condition mirrors the magnetic field vector from the lowermost domain cells into the 
boundary cells, which does result in upflow regions in a horizontal magnetic field that is organized on a scale comparable to convection cells near the bottom 
boundary, but has no preferred direction or net horizontal flux  \citep[see, ][for further detail]{Rempel:2014:SSD}. In order to highlight the role of this deep
recirculation we analyze a second model that has a similar setup as the model `Z8'  presented in \citet{Rempel:2014:SSD},
but we recomputed a 1 hour sequence with a grid spacing of 4km. This model uses an open bottom boundary that imposes $\vec{B}=0$~G in
inflow regions. The photospheric saturation field strength is around $\langle \vert B_z\vert\rangle\sim45$~G, which corresponds to 
$\langle \vert \vec{B}\vert\rangle\sim90$~G, and  $B_{\rm RMS}\sim140$~G. This model relies completely on local recirculation within 
the simulation domain, similar to \citet{Voegler:Schuessler:2007}, we refer to this model in the following as ${\rm SSD}_{\rm Shallow}$. 

\begin{figure*}
  	\centering
   	\resizebox{0.7\hscale\hsize}{!}{\includegraphics{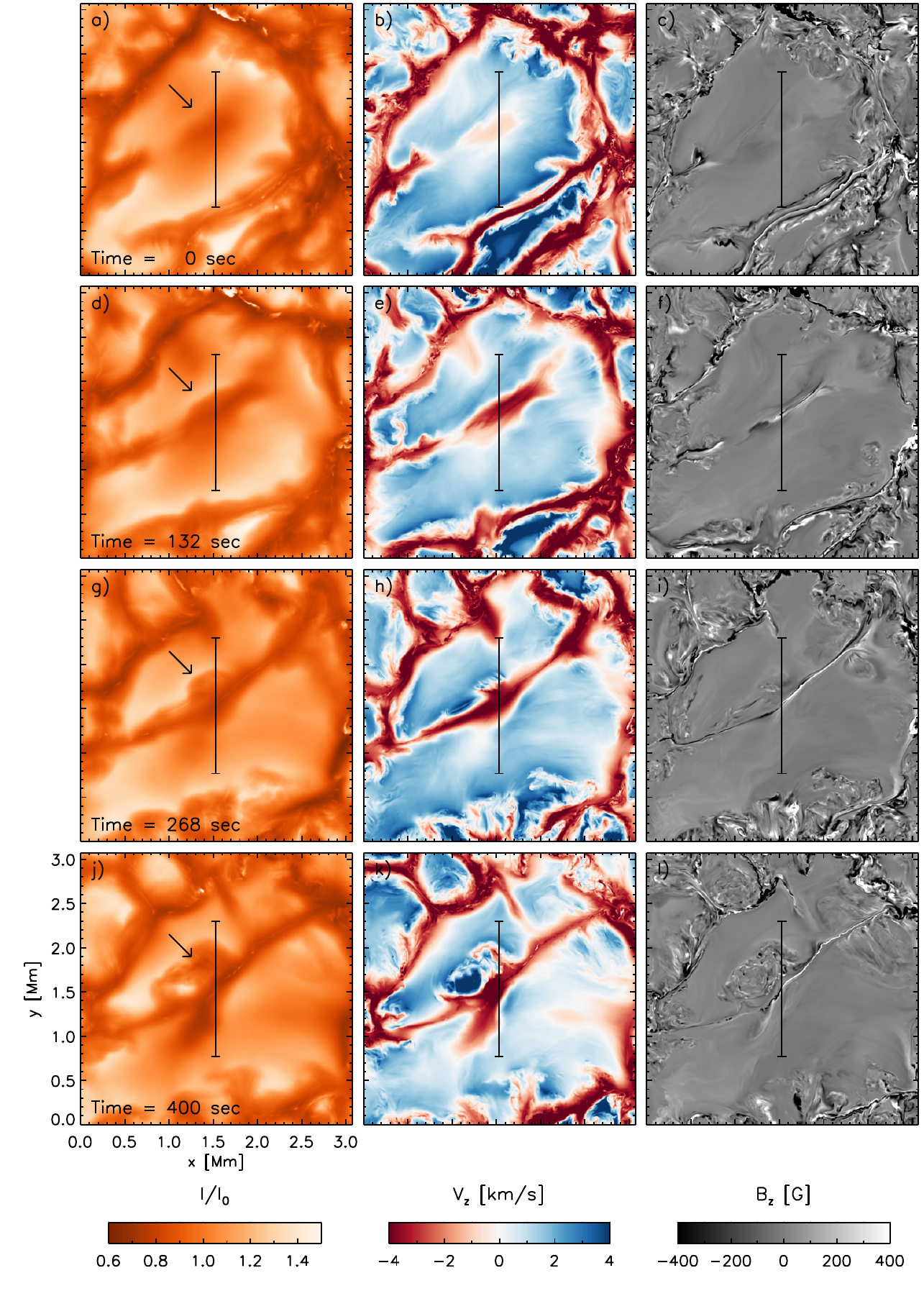}}
   	\caption{Evolution of bolometric intensity (left), vertical velocity (middle), and vertical magnetic field (right)
   		in an exploding granule over a time span of about 400 seconds (top to bottom). The vertical velocity and 
		magnetic field are extracted on the $\tau=1$ surface. The black line indicates the position of a vertical cut discussed further in Figure \ref{fig:2}. An 
		animation of this figure showing the temporal evolution over an extended time interval of 926 seconds is available.}
   	\label{fig:1}
\end{figure*}

\begin{figure*}
  	\centering
   	\resizebox{0.7\hscale\hsize}{!}{\includegraphics{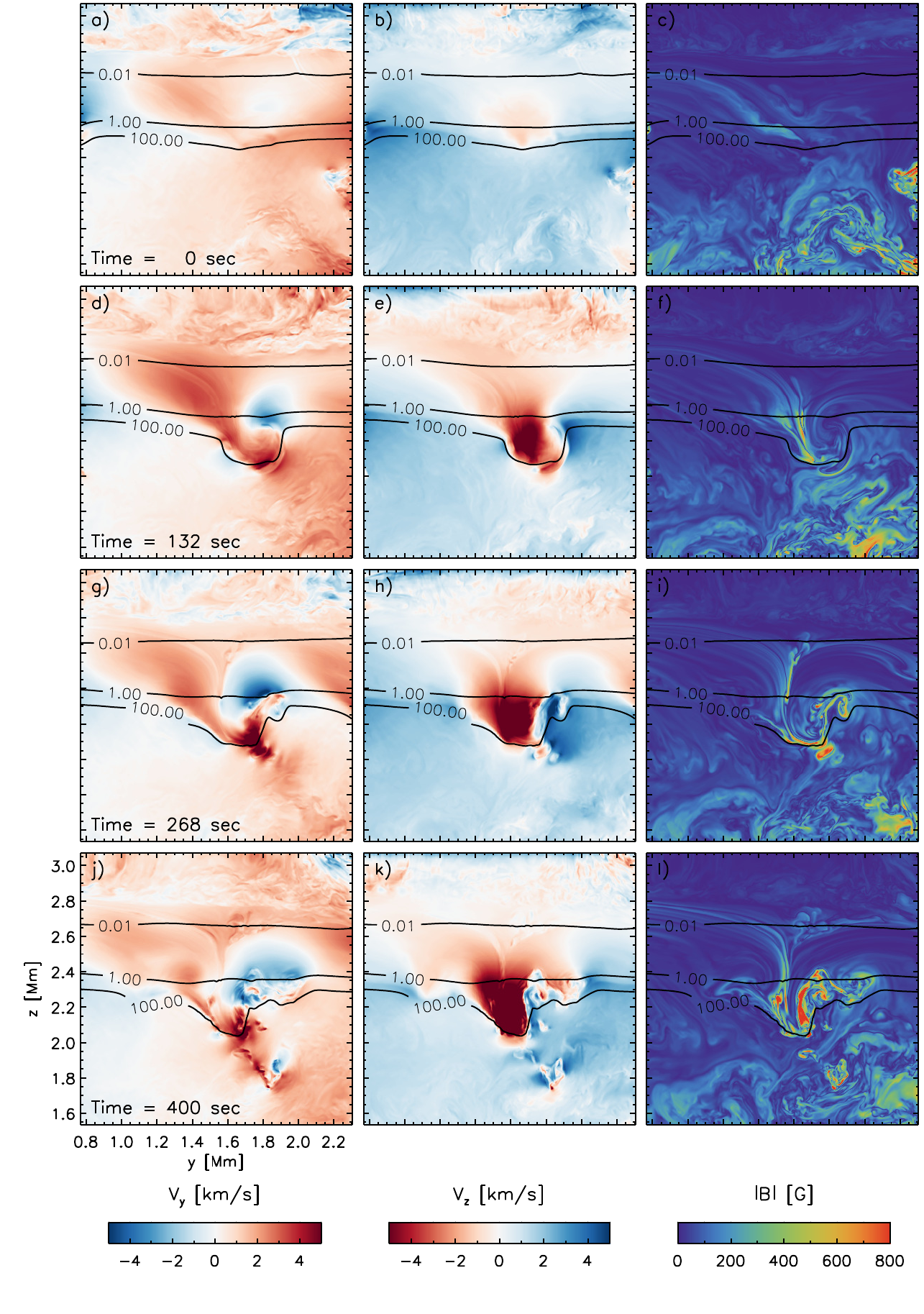}}
   	\caption{Quantities extracted along the vertical cut indicated in Figure \ref{fig:1}. Horizontal flow velocity along the cut in the
   		y-direction (left), vertical flow velocity (middle), and vertical field strength (right). We show top to bottom the same snapshots 
   		as in Figure \ref{fig:1}. Black contours indicate the $\tau$ levels of $100$, $1$, and $0.01$.}
   	\label{fig:2}
\end{figure*}

\begin{figure*}
  	\centering
   	\resizebox{0.7\hscale\hsize}{!}{\includegraphics{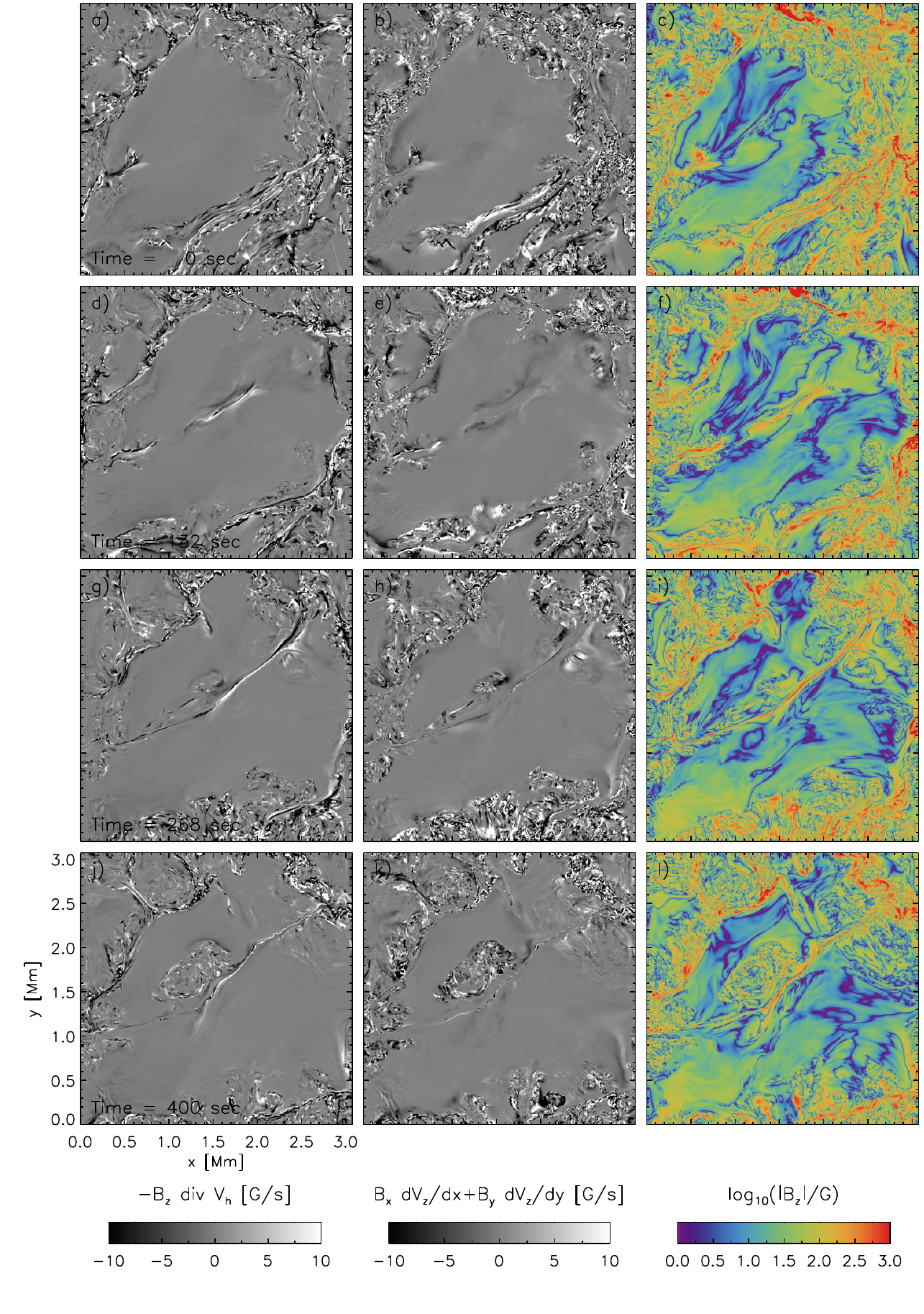}}
   	\caption{Induction terms for vertical magnetic field component from horizontal flow divergence (left) and shear (middle). The 
   		right panels show the vertical field strength on a log scale. We show top to bottom the same snapshots 
   		as in Figure \ref{fig:1}. All quantities are extracted on the $\tau=1$ surface.}
   	\label{fig:3}
\end{figure*}

\begin{figure*}
  	\centering
   	\resizebox{0.7\hscale\hsize}{!}{\includegraphics{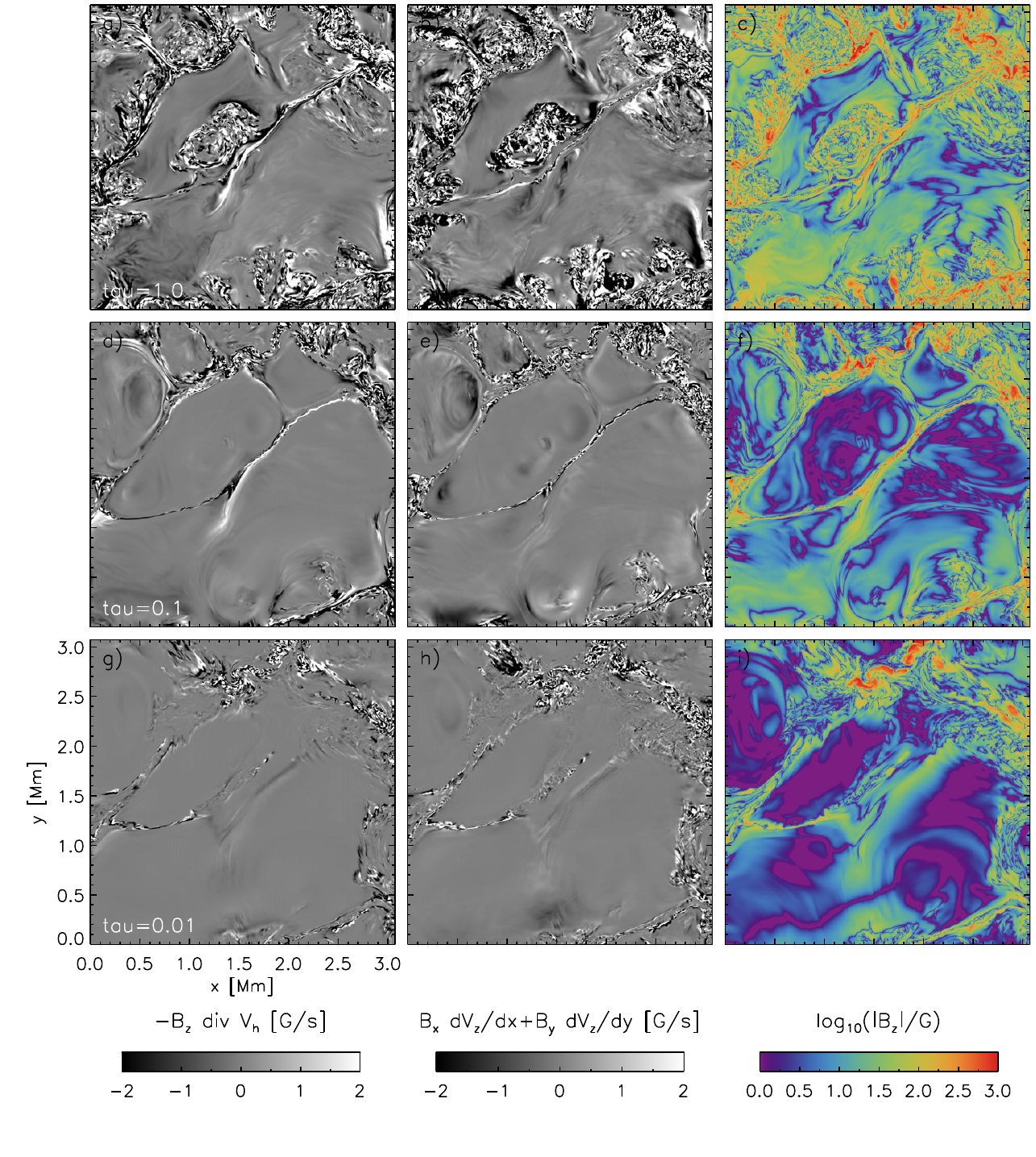}}
   	\caption{Height dependence of the induction terms shown in Figure \ref{fig:3}. We show top to bottom the $\tau$ surfaces of 
	$1$, $0.1$, and $0.01$ \NEW{for the last snapshot at $t=400$ seconds}.}
   	\label{fig:4}
\end{figure*}

\begin{figure*}
  	\centering
   	\resizebox{0.7\hscale\hsize}{!}{\includegraphics{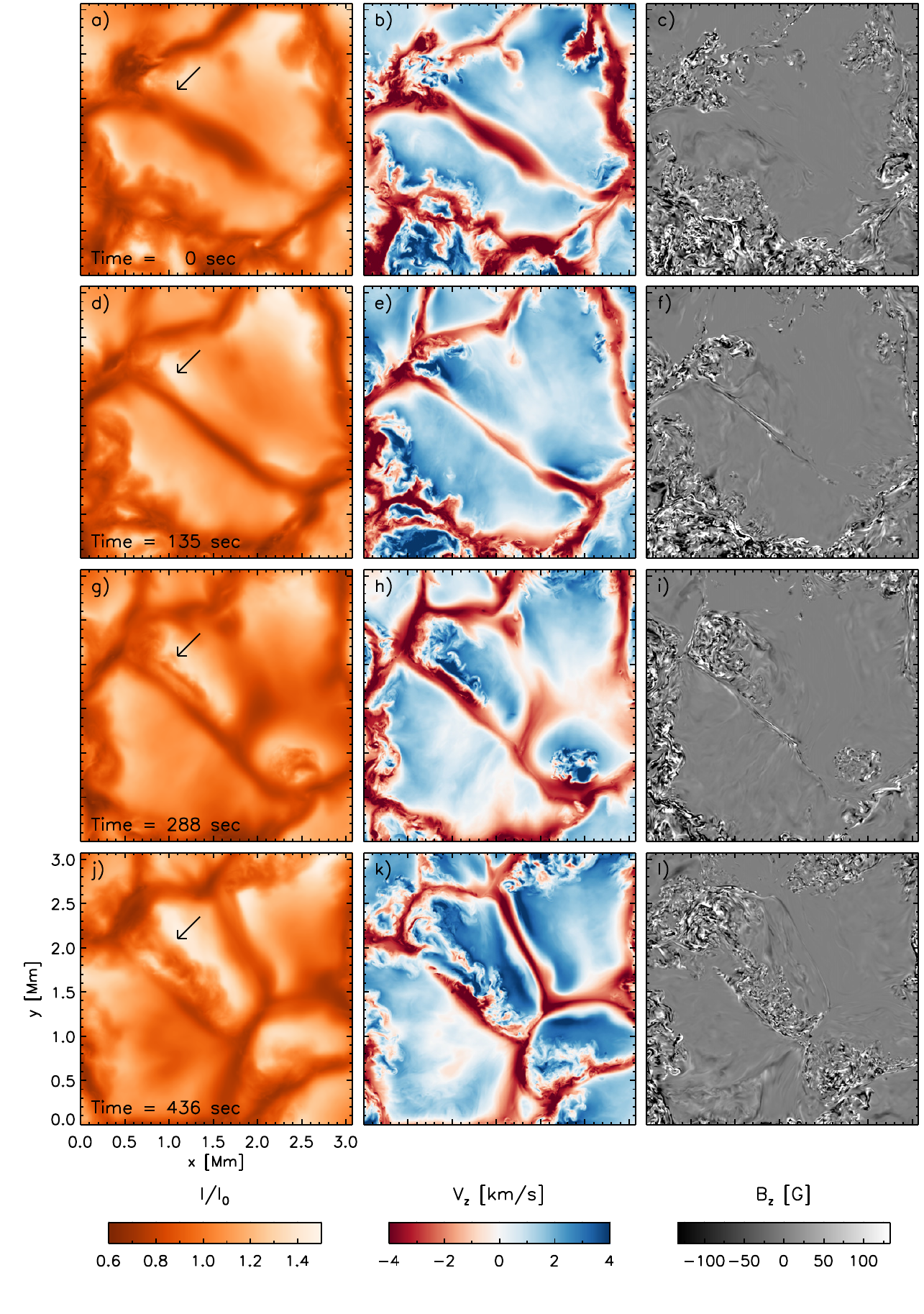}}
   	\caption{Same as Figure \ref{fig:1}, but for the simulation ${\rm SSD}_{\rm Shallow}$ ($\vec{B}=0$~G in upflows at the bottom boundary). While we do find
		a similar sequence of events as in the previous case, the overall field strength is about a factor of $3$ lower. We see less pronounced sheet-like
		magnetic structures forming in the newly developing downflow lane.}
   	\label{fig:5}
\end{figure*}

\begin{figure*}
  	\centering
   	\resizebox{0.7\hscale\hsize}{!}{\includegraphics{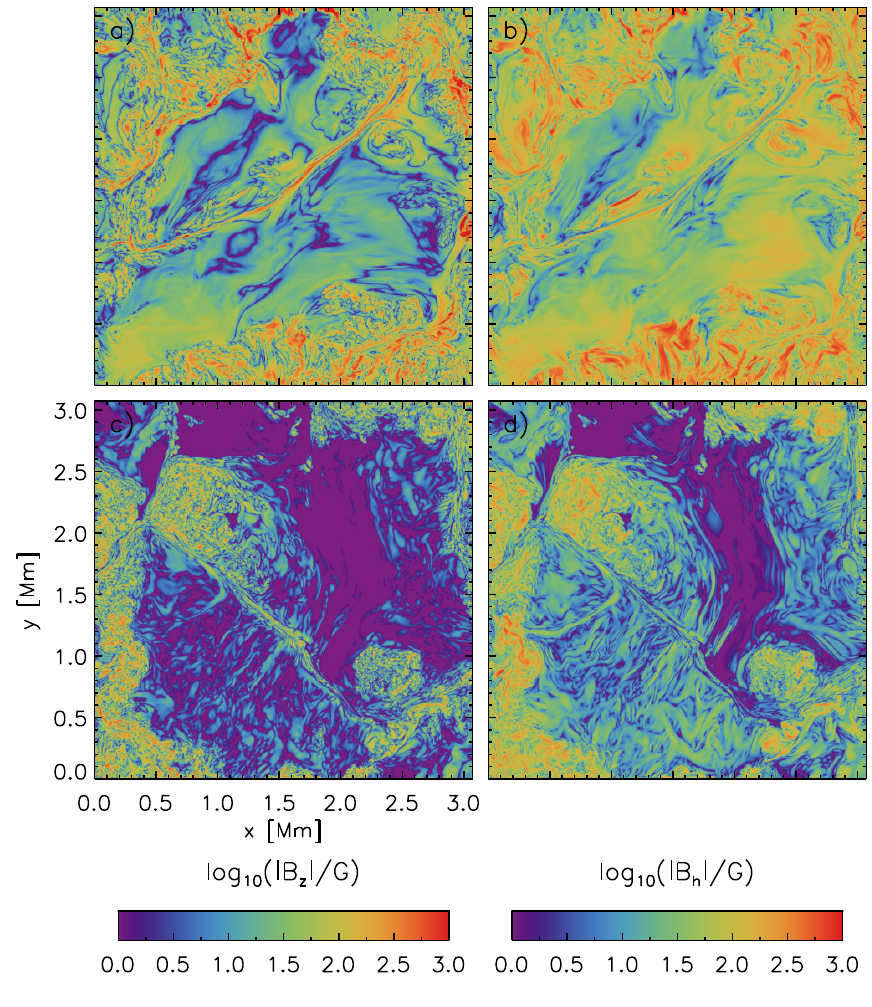}}
   	\caption{Comparison of photospheric magnetic field at $\tau=1$ in the simulation ${\rm SSD}_{\rm Deep}$ (top) and ${\rm SSD}_{\rm Shallow}$ (bottom). The panels on the left
		show $\vert B_z \vert$, the panels on the right  \NEWER{$\vert B_h \vert$}. The presented snapshots correspond to the snapshots displayed in
		Figures \ref{fig:1}g)-i) and \ref{fig:5}g)-i).}
   	\label{fig:6}
\end{figure*}

\begin{figure*}
  	\centering
   	\resizebox{0.7\hsize}{!}{\includegraphics{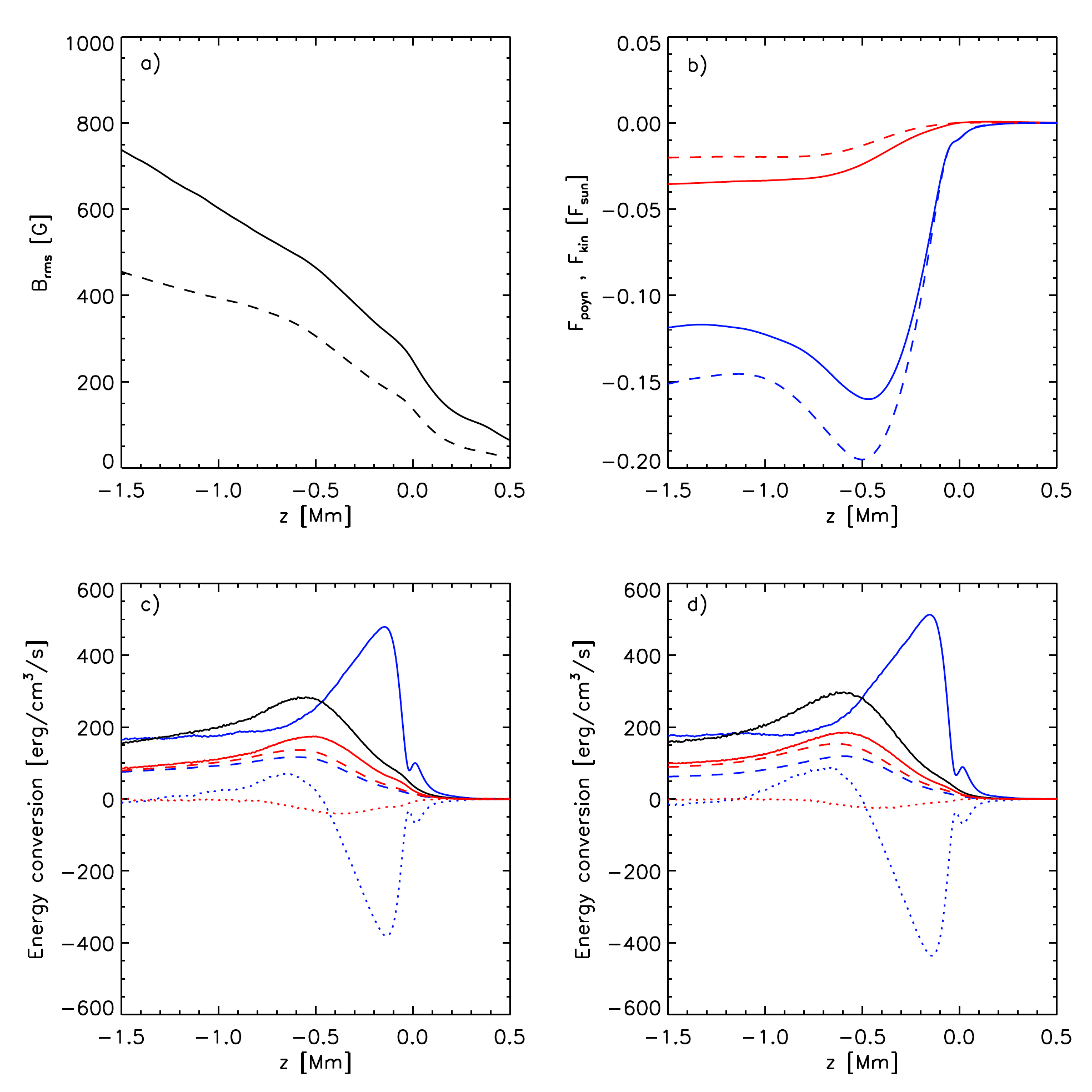}}
   	\caption{\NEW{Comparison of ${\rm SSD}_{\rm Deep}$ and ${\rm SSD}_{\rm Shallow}$: (a) RMS field strength as function of \NEWER{height ($z=0$~Mm corresponds to the average $\tau=1$ level) for} 
	${\rm SSD}_{\rm Deep}$ (solid), ${\rm SSD}_{\rm Shallow}$ (dashed). (b) Poynting flux (red) 
	and kinetic energy flux (blue) for ${\rm SSD}_{\rm Deep}$ (solid) and ${\rm SSD}_{\rm Shallow}$ (dashed). (c) Horizontally averaged energy conversion rates as function of height for ${\rm SSD}_{\rm Deep}$ as defined
	in Equations \ref{eq:1} and \ref{eq:2}: $Q_{\rm Pres/Buo}$ (blue), $Q_{\rm Kin}$ (blue, dotted) and $Q^{\rm NUM}_{visc}$ (blue, dashed), $-Q_{\rm Lorentz}$ (red) , 
	$Q_{\rm Poynting}$ (red, dotted),  $Q^{\rm NUM}_{res}$ (red, dashed). (d) Same as (c) for the case ${\rm SSD}_{\rm Shallow}$. In (c) and (d) the solid black line shows 
	$Q_{\rm Pres/Buo}+Q_{\rm Kin}$.}}
   	\label{fig:7}
\end{figure*}

\begin{figure*}
  	\centering
   	\resizebox{0.7\hsize}{!}{\includegraphics{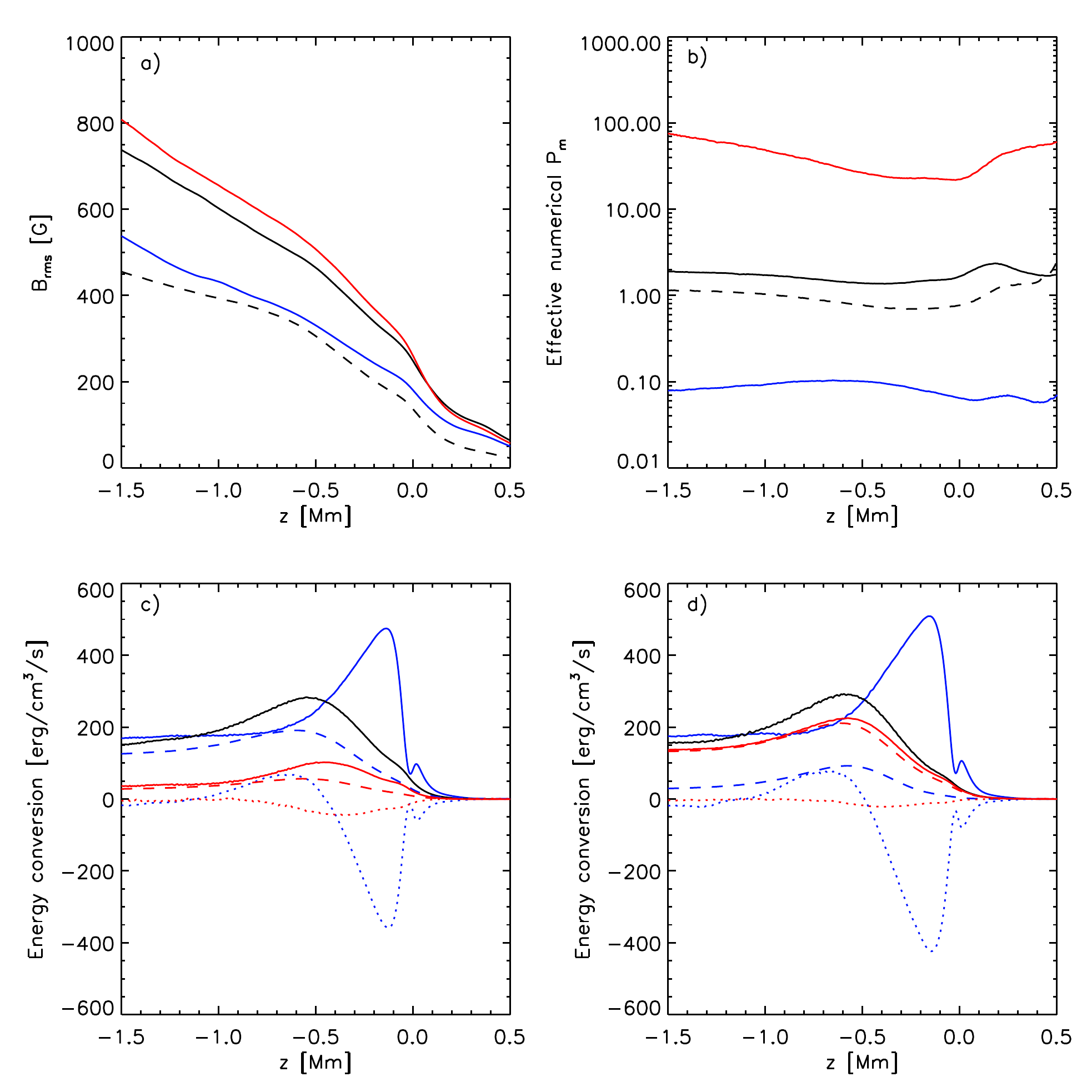}}
   	\caption{\NEW{Influence of the numerical magnetic Prandtl number, $P_{\rm m}^{\rm NUM}$, on energy conversion rates in the simulations: (a) RMS field strength as function of \NEWER{height 
	($z=0$~Mm corresponds to the average $\tau=1$ level) for} ${\rm SSD}_{\rm Deep}$ (solid), ${\rm SSD}_{\rm Shallow}$ (dashed),
	high $P_{\rm m}^{\rm NUM}$ (red),  low $P_{\rm m}^{\rm NUM}$ (blue), (b) vertical profiles of  $P_{\rm m}^{\rm NUM}$ for these cases. (c) Energy conversion terms for high $P_{\rm m}^{\rm NUM}$ case, 
	(d) Energy conversion terms for low $P_{\rm m}^{\rm NUM}$ case. We show the same quantities as in Figure \ref{fig:7}(c,d).}}
   	\label{fig:8}
\end{figure*}

\section{Results}
\subsection{Photospheric evolution}
We select in the simulation ${\rm SSD}_{\rm Deep}$, which covers during its saturated phase about 2 hours of temporal evolution, 4 exploding granules.
Figure \ref{fig:1} shows the first example, the other 3 examples can be found in the Appendix \ref{App:A}. We present top to bottom 4 snapshots spaced about 130 seconds in time (due to the
variable numerical time step the time spacing is not equidistant). Left panels show the evolution of bolometric intensity for a vertical
ray direction, middle panels the vertical velocity and right panels the vertical component of the magnetic field on the $\tau=1$ level. In the first snapshot
Figure \ref{fig:1}(a,b,c), the intensity image shows already a central darkening of the granule, the vertical velocity in the center of
the granule starts to show a weak downflow. The granule is filled with a weak vertical magnetic field of a few $10$~G strength (see
also Figure \ref{fig:3}c) that is organized on a scale comparable to the granule itself. The next snapshot Figure \ref{fig:1}(d,e,f) 132
seconds later shows now a well established downflow reaching 4 km/s in amplitude and hosting a sheet of strong vertical magnetic field.
The sheet shows a mix of polarities, since the initial field present in the granule was not uni-polar. The intensity image shows some
asymmetry of the newly formed downflow, with a brighter and sharper upper edge (indicated by arrow), which is related to a stronger 
convective upflow on that side of the granule. In the following two snapshots it is this side of the downflow lane where we find turbulent
magnetic field appearing at the edge of the adjacent granule (transported into the photosphere by the granular upflow). The turbulent field is eventually swept
into the newly formed downflow lane and leads to a turbulent magnetic field within the downflow lane with an appearance similar to that
found in well established downflow lanes.

While we focus the following analysis on the case presented in Figure \ref{fig:1}, we present in the Appendix Figures \ref{fig:9}, \ref{fig:10}, and \ref{fig:11}
three additional cases in order to highlight the robustness of the studied sequence of events.

The vertical cross-sections displayed in Figure \ref{fig:2} show more clearly the nature of the underlying flow pattern. The newly
formed downflow lane in the exploding granule remains very shallow and reaches in depth only a few 100 km beneath the $\tau=1$
level. Horizontal flows in particular in panels (d) and (g) indicate the formation of a horizontal vortex roll parallel to the downflow lane.
As first discovered by \citet{Steiner:etal:2010} such mostly horizontal vortex rolls lead to distinct features in the intensity of a granule similar to those
presented in Figure \ref{fig:1}. In addition to that the enhanced adjacent granular upflow transports magnetic field back into the photosphere
that becomes first visible at the $\tau=1$ surface in the third snapshot (panels (g)-(i) of Figures \ref{fig:1} and \ref{fig:2} ). The appearance
of this recirculated field is more turbulent in the later snapshots presented in panels (j) to (l).

Figure \ref{fig:3} presents the induction terms for the vertical magnetic field component evaluated on the $\tau=1$ surface. We compute
here the induction terms using quantities extracted on the $\tau=1$ surface rather than computing the terms on constant height surfaces
and extracting the result on the $\tau=1$ surface, since the former could be more easily accomplished based on observational data. We note that the
difference is insignificant since the  $\tau=1$ surface is not heavily distorted within the exploding granule (see Figure \ref{fig:2}). We write the induction 
equation for the vertical magnetic field component as follows:

\begin{equation}
\frac{D B_z}{D t}=\underbrace{-B_z\left(\frac{\partial v_x}{\partial x}+\frac{\partial v_y}{\partial y}\right)}_{convergence/divergence}+\underbrace{B_x\frac{\partial v_z}{\partial x}+B_y\frac{\partial v_z}{\partial y}}_{shear}
\end{equation}

Note that this decomposition of terms is different from the usual separation of compression and stretching terms, which would involve a contribution from $B_z\partial v_z/\partial z$ that
cancels out. \NEW{We prefer this decomposition since it clearly separates field amplification of existing vertical field due to horizontally convergent flows from the induction due to shear acting
on horizontal field components.}
We present the divergence/convergence and shear terms in the left and middle panels of Figure \ref{fig:3} together with the vertical field strength on a logarithmic
scale in the right panels. The convergence term dominates the early phase of the field amplification (first two snapshots), the shear term starts dominating
as soon as magnetic field with a more turbulent nature appears at the edge of the granule (last two snapshots). Overall this indicates that a newly formed
downflow lane in the photosphere remains initially mostly laminar. The granular seed field is amplified due to strong horizontal convergence of flows, which leads to induction rates exceeding
$10$~G~s$^{-1}$. Starting with a granular seed field on the order of a few $10$~G this terms produces a field exceeding $800$~G in the downflow lane at $\tau=1$ in a few minutes of time. Field amplification 
by shear (producing field with a more turbulent appearance) happens beneath the visible layers of the photosphere; however, the rather shallow recirculation transports that field back 
into the visible photosphere on a time scale of minutes.  Most of this turbulent field remains in the deep photosphere as indicated in Figure \ref{fig:4}, where we show the last snapshot of 
Figure \ref{fig:3} for the $\tau$ levels of $1$, $0.1$, and $0.01$.

In Figure \ref{fig:5} we present quantities similar to those in Figure \ref{fig:1} for the simulation ${\rm SSD}_{\rm Shallow}$ that imposes $\vec{B}=0$~G in inflow regions at the bottom boundary.
While we do find a similar series of events, the overall field strength is about a factor of $1.5-2$ lower. We find a much less pronounced sheet-like magnetic field concentration
in the newly forming downflow lane, both, in terms of strength as well as extent. \NEW{Figure \ref{fig:6} shows the vertical and horizontal field strength on the $\tau=1$ level for the simulations ${\rm SSD}_{\rm Deep}$  (top panels) and 
${\rm SSD}_{\rm Shallow}$  (bottom panels). In both cases we show the time snapshot corresponding to Figures \ref{fig:1}g)-i) and \ref{fig:5}g)-i), i.e. $t=268$ and $288$ seconds, respectively.
We find} that the less coherent sheet-like field concentrations in the simulation
${\rm SSD}_{\rm Shallow}$ are due to a much weaker and less organized seed field in the center of the granules. Whereas the simulation ${\rm SSD}_{\rm Deep}$ had a granular seed field in the $10-100$~G range that is
organized on scales comparable to the size of a granule, ${\rm SSD}_{\rm Shallow}$ shows in the central portions of the granule field strength of less than $1$ G and the granular seed field is dominated 
by small-scale field at the edge of granules that results from shallow recirculation of turbulent field. The subsequent amplification of that field leads to a weaker less coherent field in the
downflow lanes. 

\NEW{In both setups upflow regions are dominated by horizontal field, which is a consequence of horizontal expansion. The average values for $\vert B_z\vert$ ($\vert B_h\vert$) in upflow regions at $\tau=1$ are 
about 50 (100) G for ${\rm SSD}_{\rm Deep}$ and 30 (60) G for ${\rm SSD}_{\rm Shallow}$. Subsequent field amplification is more efficient for the vertical field component, leading to average values in downflow regions of 130 (160) G 
and 65 (90) G, respectively. While the downflow regions of simulation ${\rm SSD}_{\rm Shallow}$ are close to the expectation from an isotropic field distribution $\vert B_h\vert=\sqrt{2}\vert B_z\vert$, the average vertical field in ${\rm SSD}_{\rm Deep}$ is stronger 
than the expectation from an isotropic distribution due to the presence of $>1$~kG field concentrations \citep[see][for further detail]{Rempel:2014:SSD}.}  

Overall the comparison of these two models demonstrates that the magnetic field present in the center of granules plays a crucial role in maintaining the quiet sun
magnetic field. It provides the seed for further field amplification in the photosphere through a combination of concentration by convergent flows toward downflow lanes
and subsequent amplification by turbulent shear beneath the visible layers of the photosphere. The magnetic field present in the center of granules is strongly linked to magnetic
field in the deeper convection zone, which illustrates the important role for a deep seated recirculation in addition to a shallow photospheric recirculation. While the latter
appears sufficient to maintain a dynamo, the former is required to push the saturation field strength to the levels implied by observations.

\subsection{Photospheric and sub-photospheric energy conversion}
\NEW{
In this section we discuss the energy conversion rates of the small scale dynamo in the upper most $1.5$ Mm of the convection zone for the cases ${\rm SSD}_{\rm Deep}$ and ${\rm SSD}_{\rm Shallow}$. To this end
we focus on the kinetic and magnetic energy equations:
\begin{eqnarray}
	\frac{\partial}{\partial t}\left(\frac{1}{2}\varrho v^2\right) &=&\underbrace{ -\nabla\cdot\left(\vec{v} \frac{1}{2}\varrho v^2\right)}_{Q_{\rm Kin}}\underbrace{-\vec{v}\cdot(\nabla p - \varrho\vec{g})}_{Q_{\rm Pres/Buo}}+\underbrace{\vec{v}\cdot\frac{(\nabla\times\vec{B})\times\vec{B}}{4\pi}}_{Q_{\rm Lorentz}}+\vec{v}\cdot \vec{d}^{\rm NUM}_v\label{eq:1}\\
	\frac{\partial}{\partial t}\left(\frac{B^2}{8\pi}\right) &=& \underbrace{-\nabla\cdot\left(\frac{\vec{B}\times(\vec{v}\times\vec{B})}{4\pi}\right)}_{Q_{\rm Poynting}}-\vec{v}\cdot\frac{(\nabla\times\vec{B})\times\vec{B}}{4\pi}+\frac{1}{4\pi}\vec{B}\cdot \vec{d}^{\rm NUM}_B\label{eq:2}
\end{eqnarray}
Here $\varrho$, $p$, $\vec{v}$, and $\vec{B}$ denote mass density, pressure, velocity and magnetic field. $\vec{d}^{\rm NUM}_v$ and $\vec{d}^{\rm NUM}_B$ indicate numerical diffusion terms in the momentum and induction equation.
The work by the numerical diffusion terms $\vec{v}\cdot \vec{d}^{\rm NUM}_v$ and $\vec{B}\cdot \vec{d}^{\rm NUM}_B/4\pi$ can be expressed as
\begin{eqnarray}
	\vec{v}\cdot \vec{d}^{\rm NUM}_v &=& -\nabla\cdot [...]-Q^{\rm NUM}_{visc}\label{eq:3}\\
	\frac{1}{4\pi}\vec{B}\cdot \vec{d}^{\rm NUM}_B&=& -\nabla\cdot [...]-Q^{\rm NUM}_{res}\label{eq:4}
\end{eqnarray}
where the first terms on the right hand side of Equations (\ref{eq:3}) and (\ref{eq:4}) indicate viscous and resistive energy fluxes that turn out to be insignificant for the numerical diffusivities used.
We use in the following discussion $Q^{\rm NUM}_{visc}$ and $Q^{\rm NUM}_{res}$ that are computed as specified in the Appendix \ref{App:B} \citep[see, also][]{Rempel:2014:SSD,Rempel:2017:corona,Hotta:2017:Dissipation}.}

\NEW{
Figure \ref{fig:7}a) compares the vertical profiles of the RMS field strength in the runs ${\rm SSD}_{\rm Deep}$ (solid) and ${\rm SSD}_{\rm Shallow}$ (dashed). ${\rm SSD}_{\rm Deep}$ has throughout the domain a field strength that is about a factor of
$1.5-2$ stronger than ${\rm SSD}_{\rm Deep}$ depending on the position. Figure \ref{fig:7}b) compares the Poynting (red) and kinetic energy (blue) fluxes for both models. Although ${\rm SSD}_{\rm Deep}$ has a bottom boundary that
mimics a deep convection zone by allowing transport of horizontal field into the domain in upflow regions (upward Poynting flux), that contribution is overcompensated  by contributions from downflow regions, leading to an overall stronger 
downward directed Poynting flux. At the same time Lorentz force feedback reduces the amplitude of the downward directed 
kinetic energy flux, which overcompensates the contribution from the Poynting flux. These trends become even more pronounced in deeper reaching models as discussed in \citet{Hotta:2015:SSD}.}

\NEW{
Figure \ref{fig:7}c) presents an analysis of the energy conversion terms as defined in Equations (\ref{eq:1}) to  (\ref{eq:4}) for the run ${\rm SSD}_{\rm Deep}$. For this analysis we restarted the simulation and extracted
horizontal averages of the respective terms at a high time cadence (about every 5-10 seconds) while the simulation was running for a simulated time span of about 1 hour. In particular the numerical dissipation terms $Q^{\rm NUM}_{visc}$ and 
$Q^{\rm NUM}_{res}$ need to be extracted directly from then simulation while it is running and cannot be computed in a post processing step (applying a filter to an already filtered solution would provide a 
different result). We provide the expressions for $Q^{\rm NUM}_{visc}$ and  $Q^{\rm NUM}_{res}$ in the Appendix \ref{App:B}. The primary driver (energy source) for convective motions is pressure buoyancy driving ($Q_{\rm Pres/Buo}$, blue, solid)
that peaks about $100$~km beneath $\tau=1$. Most of that driving leads to a strong downward acceleration of fluid, evident in a substantial kinetic energy flux as shown in Figure \ref{fig:7}b). The negative divergence of the kinetic energy flux, $Q_{\rm Kin}$, is shown in 
Figure \ref{fig:7}c) as blue dotted line. The sum of $Q_{\rm Pres/Buo}$ and $Q_{\rm Kin}$ (black, solid) indicates as function of \NEWER{height} the power that is available for sustaining  turbulence against viscous
dissipation as well as the small scale dynamo. This quantity peaks in a depth of $500$ to $600$~km beneath the photosphere. Averaged between $z=-1.5$~Mm and $z=0$~Mm about $59\,\%$ of 
$Q_{\rm Pres/Buo}+Q_{\rm Kin}$ are transferred via Lorentz force work $Q_{\rm Lorentz}$ to magnetic energy. We show the quantity  $-Q_{\rm Lorentz}$ (red, solid), which is the primary energy source in the
magnetic energy equation. $13\,\%$ of that transfer goes into the divergence of the Poynting flux,
$Q_{\rm Poynting}$ (red, dotted),  and the remainder into (numerical) resistive dissipation, $Q^{\rm NUM}_{res}$ (red, dashed), which has a similar magnitude as the (numerical) viscous dissipation, 
$Q^{\rm NUM}_{visc}$ (blue, dashed). Figure \ref{fig:7}d) shows the same terms for the run ${\rm SSD}_{\rm Shallow}$. Although that simulation maintains magnetic energy at an about 3 times lower level, \NEWER{the overall dynamo energetics 
are very similar to those of ${\rm SSD}_{\rm Deep}$.} In order to better understand this result, we look next into the role of the effective numerical magnetic Prandtl number.}

\NEW{
We follow here \citet{Rempel:2017:corona} to estimate effective numerical viscosity and resistivity by comparing the numerical viscous and resistive energy dissipation rates $Q^{\rm NUM}_{visc}$ and $Q^{\rm NUM}_{res}$ 
to the equivalent rates for explicit diffusivities:
\begin{eqnarray}
        \epsilon_{\nu}&=&\nu\varrho \sum_{i,k}\frac{\partial v_i}{\partial x_k}\left[\frac{\partial v_i}{\partial x_k}+\frac{\partial v_k}{\partial x_i}-\frac{2}{3}\delta_{ik}\nabla\cdot\vec{v}\right]\\
        \epsilon_{\eta}&=&\frac{\eta}{4\pi} \vert\nabla\times\vec{B}\vert^2
\end{eqnarray}
Since numerical diffusivities are strongly intermittent and can be even zero depending on the local smoothness of the solution, a point by point comparison is not necessarily meaningful. We consider instead quantities averaged over horizontal planes, indicated by $\langle \ldots \rangle$, leading to
the effective numerical diffusivities:
  \begin{eqnarray}
        \nu_{\rm eff}&=&\frac{\langle Q^{\rm NUM}_{visc} \rangle}{ \langle \varrho \sum_{i,k}\frac{\partial v_i}{\partial x_k}\left[\frac{\partial v_i}{\partial x_k}+\frac{\partial v_k}{\partial x_i}-\frac{2}{3}\delta_{ik}\nabla\cdot\vec{v}\right]\rangle}\\
        \eta_{\rm eff}&=&4\pi \frac{\langle Q^{\rm NUM}_{res} \rangle}{ \langle \vert\nabla\times\vec{B}\vert^2 \rangle}
\end{eqnarray}
from which we can compute the effective numerical magnetic Prandtl number as $P_m^{\rm NUM}= \nu_{\rm eff}/\eta_{\rm eff}$ as function of height.}

\NEW{
In addition to the runs ${\rm SSD}_{\rm Deep}$ and ${\rm SSD}_{\rm Shallow}$ we computed for the ${\rm SSD}_{\rm Deep}$ setup also a high and a low $P_m^{\rm NUM}$ case by combining different diffusivity settings for numerical viscosity and 
resistivity. As described in \citet{Rempel:2014:SSD,Rempel:2017:corona} and the Appendix \ref{App:B}, the formulation of numerical diffusivities has a parameter $h$ that determines how concentrated numerical diffusivities are near monotonicity changes.
While we use in our standard setting $h=2$ for all variables, we emulate a high $P_m^{\rm NUM}$ case through $h=(0, 4)$ for $(\vec{v}, \vec{B})$, and a low $P_m^{\rm NUM}$ case through $h=(4, 0)$. Note that this approach does not keep the magnetic Reynolds number fixed, i.e. the low $P_m^{\rm NUM}$ case has also the lowest value of $R_{\rm m}^{\rm NUM}$. The average numerical magnetic resistivity in the low $P_m^{\rm NUM}$ case is with $1.6\cdot 10^{10}$cm$^2$ s$^{-1}$ about $26$ times larger than in the
high $P_m^{\rm NUM}$ case with $6.1\cdot 10^{8}$cm$^2$ s$^{-1}$. Using a typical granular size of 1~Mm as length scale and a granular RMS velocity of 3~km/s as velocity scale, the corresponding $R_{\rm m}^{\rm NUM}$ values are about $2,000$
and $50,000$, respectively. The former is close to the critical $R_{\rm m}^{\rm NUM}$ found in \citet{Voegler:Schuessler:2007}.}

\NEW{
Figure \ref{fig:8}a) shows the vertical RMS field strength for the high (red) and low (blue) $P_m^{\rm NUM}$ together with those for the ${\rm SSD}_{\rm Deep}$ and ${\rm SSD}_{\rm Shallow}$ cases from Figure \ref{fig:7}a). The low $P_m^{\rm NUM}$ 
case has a saturation field strength almost comparable to ${\rm SSD}_{\rm Shallow}$, which is due to an overall lower value of $R_{\rm m}^{\rm NUM}$. Figure \ref{fig:8}b) presents the $P_m^{\rm NUM}$ values for these cases. 
Depending on the location, the high $P_m^{\rm NUM}$ has values larger than $20$, the cases ${\rm SSD}_{\rm Deep}$ and ${\rm SSD}_{\rm Shallow}$ between 1 and 2, and the low $P_m^{\rm NUM}$ case around $0.1$. }

\NEW{
Figure \ref{fig:8}c) and \ref{fig:8}d) show the energy conversion terms for the high and low  $P_m^{\rm NUM}$  case, respectively. As discussed by \citet{Brandenburg:2011:SSD_low_Pm,Brandenburg:2014:Pm} we find
also here with purely numerical diffusivities that $P_m^{\rm NUM}$ controls the ratio between $Q^{\rm NUM}_{visc}$ and $Q^{\rm NUM}_{res}$, which is about $3$ for the high $P_m^{\rm NUM}$  and
about $0.3$ for the low $P_m^{\rm NUM}$ case, note that the latter has at the same time the lower saturation field strength. Although the cases ${\rm SSD}_{\rm Deep}$ and ${\rm SSD}_{\rm Shallow}$ use the same numerical diffusivity settings,
they differ in their value of $P_m^{\rm NUM}$ by about a factor of $2$, with ${\rm SSD}_{\rm Shallow}$ having the lower value. 
${\rm SSD}_{\rm Shallow}$ has a slightly lower $P_m^{\rm NUM}$ as ${\rm SSD}_{\rm Deep}$ since magnetic field in ${\rm SSD}_{\rm Shallow}$ is organized on smaller scales than in ${\rm SSD}_{\rm Deep}$
and is consequently more strongly affected by numerical dissipation terms.}
\NEWER{The similar energy conversion rates in ${\rm SSD}_{\rm Deep}$ and  ${\rm SSD}_{\rm Shallow}$ (Figure \ref{fig:7}) are a consequence of comparable effective numerical magnetic Prandtl numbers within a factor of $2$.}

\NEW{
Overall our findings are consistent (at least on a qualitative level) with those reported by  \citet{Brandenburg:2011:SSD_low_Pm,Brandenburg:2014:Pm} using simulations of forced turbulence with explicit viscosity and magnetic diffusivity. For our lowest value
of $P_m^{\rm NUM}\sim 0.1 $ we find that $Q_{\rm Lorentz}$ is about $80\,\%$ of $Q_{\rm Pres/Buo}+Q_{\rm Kin}$, whereas it is only $30\,\%$ in the high $P_m^{\rm NUM}$ case. \NEWER{Assuming that this scaling extends to $P_{\rm m}$ values as low as 
$10^{-5}$ found in the solar photosphere,} it is conceivable that  $Q_{\rm Lorentz}$ would reach close to $100\,\%$ of $Q_{\rm Pres/Buo}+Q_{\rm Kin}$, i.e. Maxwell-stresses take over the role
of Reynolds-stresses and viscous dissipation becomes insignificant. Integrated over the uppermost $1.5$ Mm of the convection zone $Q_{\rm Pres/Buo}+Q_{\rm Kin}$ reaches \NEWER{in the simulation ${\rm SSD}_{\rm Deep}$} a value of $46\,\%$ of the
solar luminosity \NEWER{(the other cases considered are within a few $\%$ of this value)}, i.e. the energy conversion by the small-scale dynamo is expected to be quite substantial.
}
 
\section{Conclusions}
We analyzed the magnetic field amplification in exploding granules and focused in particular on the time evolution of newly formed downflow lanes.
We found the following sequence of events in at least four exploding granules present in the analyzed simulation run ${\rm SSD}_{\rm Deep}$:
\begin{enumerate}
	\item Horizontal flows converging towards the newly forming downflow lane amplify a weak (a few $10$~G) field present in granular upflows to
		a strength exceeding $800$~G within a time span of a few minutes.
		The newly forming downflow lane remains mostly laminar, i.e. field amplification due to turbulent velocity shear remains weak.
	\item The newly forming downflow lane develops asymmetric horizontal vorticity, which is manifest early on in the intensity image in form
		of sharp intensity gradients on one side of the downflow lane \citep{Steiner:etal:2010}.
	\item Within a few minutes magnetic field with a more turbulent appearance becomes visible at the edge of the adjacent granule that showed previously 
	        the shaper intensity gradients. The turbulent magnetic field is swept into the downflow lane and leads to the presence of turbulent field (and flows) in 
	        the newly formed downflow lane.	
\end{enumerate}

\NEW{In the early stages of this process vertical magnetic field is primarily amplified by horizontally converging}
flows, which is an unavoidable consequence of overturning granular motions. This term is strong enough to produce narrow sheets of vertical field reaching
more than $800$~G in a few minutes. The structure of these sheets in terms of extent along the downflow lane as well as polarity (some sheets are unipolar,
whereas others may have mixed polarity) is a reflection of the structure in the granular seed field. Since plasma that appears in the center of granules has undergone
substantial horizontal expansion, magnetic field in the center of granules tends to be organized on scales comparable to the granular extent. Thus studying the structure
of magnetic field in newly forming downflow lanes provides a way to quantify the strength and structure of magnetic field that reaches the photosphere from the deeper convection
zone. This deep recirculation component provides a significant contribution to small-scale magnetism in the photosphere. \NEW{The simulation ${\rm SSD}_{\rm Shallow}$ that assumes $\vec{B}=0$~G
in upflows at the bottom boundary has an about $1.5-2$ times lower saturation level. Such values falls short in 
comparison with observations \citep{Danilovic:2010:zeeman_dynamo,Danilovic:2016:SSD_Power}. In addition to that the sheet-like organization of magnetic field is less
pronounced.} 

We did focus our discussion mostly on exploding granules since they provide the most undisturbed view on the field amplification process, the described picture is however
representative for the photosphere as a whole. 

We note that current simulations might not properly capture the structure of the deep recirculation component. 
On the one hand, the finite domain depth underestimates the smoothing due to horizontal expansion; on the other hand, the finite grid spacing imposes a lower limit to the scale 
of magnetic field at the bottom boundary, which becomes amplified due to horizontal expansion. Addressing this problem in simulations requires the combination of deeper domains 
with higher resolution, which is numerically expensive. Therefore observational constraints from current and future high resolution telescopes are required for further progress. 

\NEW{ Interestingly, the total energy conversion associated with small-scale dynamo action is not affected by the presence or absence of deep recirculation. We find in both ${\rm SSD}_{\rm Deep}$ and ${\rm SSD}_{\rm Shallow}$ comparable
amounts of work against the Lorentz force in spite of different saturation field strengths. As previously discovered by \citet{Brandenburg:2011:SSD_low_Pm,Brandenburg:2014:Pm} we also find that the magnetic 
Prandtl number (in our case the effective numerical magnetic Prandtl number,  $P_m^{\rm NUM}$) determines the overall energy conversion through the Lorentz force/induction terms and consequently the ratio of
viscous to resistive energy conversion. For a value of $P_m^{\rm NUM}\sim 0.1$, $Q_{\rm Lorentz}$ is about $80\,\%$ of $Q_{\rm Pres/Buo}+Q_{\rm Kin}$ (local convective driving), which translates to more than $30\,\%$ of the
solar luminosity when integrated over the uppermost $1.5$~Mm of the convection zone. \NEWER{Assuming that this scaling extends to $P_{\rm m}$ values as low as the $10^{-5}$ found in the solar photosphere,} $Q_{\rm Lorentz}$  could be 
close to $100\,\%$ of $Q_{\rm Pres/Buo}+Q_{\rm Kin}$. 
\NEWER{In the presented simulations} Maxwell-stresses are of substantial strength and take over the role of Reynolds stresses and viscous dissipation. While less clear in the near surface layers, this does influence the structure
of convection of turbulence in deeper simulations domains as demonstrated by \citet{Hotta:2015:SSD}. They compare pure hydrodynamic and small-scale dynamo simulations and found in the latter a reduction of the upflow/downflow mixing somewhat
similar to high thermal Prandtl number convection, which is expected at least on a qualitative level when the Maxwell-stress mimics the effect of turbulent stresses. 
}

\NEW{We do not find strong {\it direct} evidence for turbulent field amplification in the {\it visible} layers of the photosphere. Turbulent field appears first in granular upflows near the edge of granules, indicating subsurface field amplification and
shallow recirculation.} 
Overall our analysis shows that small-scale field in the visible layers of the photosphere has two distinct contributions that can be linked to shallow and deep recirculation. Shallow recirculation provides the
primary source for field with turbulent appearance on small scales, whereas deep recirculation can be linked to strong sheet-like field structures in downflow lanes that do have a coherence
comparable to the extent of granules. Models with only shallow recirculation (${\rm SSD}_{\rm Shallow}$) fall short by about a factor of $1.5-2$ compared to observations, whereas models with deep recirculation (${\rm SSD}_{\rm Deep}$) do 
match observational constraints. High resolution observations of the deep photosphere could provide valuable constraints on the relative contribution from these two processes. \NEW{In both cases the 
total energy conversion by Maxwell stresses is substantial.} 

\acknowledgements
The National Center for Atmospheric Research (NCAR) is sponsored by the National Science Foundation. The author thanks Valentin Martinez-Pillet, Rebecca Centeno-Elliott and the anonymous referee for valuable comments 
on the manuscript. High-performance computing resources were provided by NCAR's Computational and Information Systems Laboratory, sponsored by the National Science Foundation, on 
Yellowstone (http://n2t.net/ark:/85065/d7wd3xhc) and Cheyenne (doi:10.5065/D6RX99HX). 

\bibliographystyle{aasjournal}
\bibliography{natbib/papref_m}

\begin{thebibliography}{}
\expandafter\ifx\csname natexlab\endcsname\relax\def\natexlab#1{#1}\fi

\bibitem[{{Boldyrev} \& {Cattaneo}(2004)}]{Bouldurev:Cattaneo:2004}
{Boldyrev}, S., \& {Cattaneo}, F. 2004, Physical Review Letters, 92, 144501

\bibitem[{{Brandenburg}(2011)}]{Brandenburg:2011:SSD_low_Pm}
{Brandenburg}, A. 2011, \apj, 741, 92

\bibitem[{{Brandenburg}(2014)}]{Brandenburg:2014:Pm}
---. 2014, \apj, 791, 12

\bibitem[{{Buehler} {et~al.}(2013){Buehler}, {Lagg}, \&
  {Solanki}}]{Buehler:2013:cyc_dep}
{Buehler}, D., {Lagg}, A., \& {Solanki}, S.~K. 2013, \aap, 555, A33

\bibitem[{{Bushby} \& {Favier}(2014)}]{Bushby:Favier:2014}
{Bushby}, P.~J., \& {Favier}, B. 2014, \aap, 562, A72

\bibitem[{{Cattaneo}(1999)}]{Cattaneo:1999}
{Cattaneo}, F. 1999, \apj, 515, L39

\bibitem[{{Danilovic} {et~al.}(2016){Danilovic}, {Rempel}, {van Noort}, \&
  {Cameron}}]{Danilovic:2016:SSD_Power}
{Danilovic}, S., {Rempel}, M., {van Noort}, M., \& {Cameron}, R. 2016, \aap,
  594, A103

\bibitem[{{Danilovic} {et~al.}(2010){Danilovic}, {Sch{\"u}ssler}, \&
  {Solanki}}]{Danilovic:2010:zeeman_dynamo}
{Danilovic}, S., {Sch{\"u}ssler}, M., \& {Solanki}, S.~K. 2010, \aap, 513, A1

\bibitem[{{Hotta}(2017)}]{Hotta:2017:Dissipation}
{Hotta}, H. 2017, \apj, 845, 164

\bibitem[{{Hotta} {et~al.}(2015){Hotta}, {Rempel}, \&
  {Yokoyama}}]{Hotta:2015:SSD}
{Hotta}, H., {Rempel}, M., \& {Yokoyama}, T. 2015, \apj, 803, 42

\bibitem[{{Ishikawa} \& {Tsuneta}(2009)}]{Ishikawa:2009:cmp_QS_plage}
{Ishikawa}, R., \& {Tsuneta}, S. 2009, \aap, 495, 607

\bibitem[{{Iskakov} {et~al.}(2007){Iskakov}, {Schekochihin}, {Cowley},
  {McWilliams}, \& {Proctor}}]{Iskakov:2007:SSD_low_Pm}
{Iskakov}, A.~B., {Schekochihin}, A.~A., {Cowley}, S.~C., {McWilliams}, J.~C.,
  \& {Proctor}, M.~R.~E. 2007, Physical Review Letters, 98, 208501

\bibitem[{{Khomenko} {et~al.}(2017){Khomenko}, {Vitas}, {Collados}, \& {de
  Vicente}}]{Khomenko:2017:Biermann}
{Khomenko}, E., {Vitas}, N., {Collados}, M., \& {de Vicente}, A. 2017, \aap,
  604, A66

\bibitem[{{Kitiashvili} {et~al.}(2015){Kitiashvili}, {Kosovichev}, {Mansour},
  \& {Wray}}]{Kitiashvili:2015:SSD}
{Kitiashvili}, I.~N., {Kosovichev}, A.~G., {Mansour}, N.~N., \& {Wray}, A.~A.
  2015, \apj, 809, 84

\bibitem[{{Lamb} {et~al.}(2014){Lamb}, {Howard}, \&
  {DeForest}}]{Lamb:etal:2014}
{Lamb}, D.~A., {Howard}, T.~A., \& {DeForest}, C.~E. 2014, \apj, 788, 7

\bibitem[{{Lites}(2011)}]{Lites:2011:hinode_ssd}
{Lites}, B.~W. 2011, \apj, 737, 52

\bibitem[{{Lites} {et~al.}(2014){Lites}, {Centeno}, \&
  {McIntosh}}]{Lites:2014:cycle_dep}
{Lites}, B.~W., {Centeno}, R., \& {McIntosh}, S.~W. 2014, \pasj, 66, S4

\bibitem[{{Petrovay} \& {Szakaly}(1993)}]{Petrovay:1993:SSD}
{Petrovay}, K., \& {Szakaly}, G. 1993, \aap, 274, 543

\bibitem[{{Pietarila Graham} {et~al.}(2010){Pietarila Graham}, {Cameron}, \&
  {Sch{\"u}ssler}}]{Pietarila-Graham:etal:2010:SSD}
{Pietarila Graham}, J., {Cameron}, R., \& {Sch{\"u}ssler}, M. 2010, \apj, 714,
  1606

\bibitem[{{Rempel}(2014)}]{Rempel:2014:SSD}
{Rempel}, M. 2014, \apj, 789, 132

\bibitem[{{Rempel}(2017)}]{Rempel:2017:corona}
---. 2017, \apj, 834, 10

\bibitem[{{Schekochihin} {et~al.}(2005){Schekochihin}, {Haugen}, {Brandenburg},
  {Cowley}, {Maron}, \& {McWilliams}}]{Schekochihin:etal:2005}
{Schekochihin}, A.~A., {Haugen}, N.~E.~L., {Brandenburg}, A., {et~al.} 2005,
  \apjl, 625, L115

\bibitem[{{Schekochihin} {et~al.}(2007){Schekochihin}, {Iskakov}, {Cowley},
  {McWilliams}, {Proctor}, \& {Yousef}}]{Schekochihin:etal:2007}
{Schekochihin}, A.~A., {Iskakov}, A.~B., {Cowley}, S.~C., {et~al.} 2007, New
  Journal of Physics, 9, 300

\bibitem[{{Stein} {et~al.}(2003){Stein}, {Bercik}, \&
  {Nordlund}}]{Stein:2003:review}
{Stein}, R.~F., {Bercik}, D., \& {Nordlund}, {\AA}. 2003, in Astronomical
  Society of the Pacific Conference Series, Vol. 286, Current Theoretical
  Models and Future High Resolution Solar Observations: Preparing for ATST, ed.
  A.~A. {Pevtsov} \& H.~{Uitenbroek}, 121

\bibitem[{{Steiner} {et~al.}(2010){Steiner}, {Franz}, {Bello Gonz{\'a}lez},
  {Nutto}, {Rezaei}, {Mart{\'{\i}}nez Pillet}, {Bonet Navarro}, {del Toro
  Iniesta}, {Domingo}, {Solanki}, {Kn{\"o}lker}, {Schmidt}, {Barthol}, \&
  {Gandorfer}}]{Steiner:etal:2010}
{Steiner}, O., {Franz}, M., {Bello Gonz{\'a}lez}, N., {et~al.} 2010, \apjl,
  723, L180

\bibitem[{{Thaler} \& {Spruit}(2015)}]{Thaler:Spruit:2015:SSD_Pm}
{Thaler}, I., \& {Spruit}, H.~C. 2015, \aap, 578, A54

\bibitem[{{Tobias} {et~al.}(2013){Tobias}, {Cattaneo}, \&
  {Boldyrev}}]{Tobias:Cattaneao:Boldyrev:SSD_review}
{Tobias}, S.~M., {Cattaneo}, F., \& {Boldyrev}, S. 2013, in {Ten Chapters in
  Turbulence}, ed. P.~A. {Davidson}, Y.~{Kaneda}, \& K.~R. {Sreenivasan}
  (Cambridge: Cambridge Univ. Press), 351--404

\bibitem[{{V{\"o}gler} \& {Sch{\"u}ssler}(2007)}]{Voegler:Schuessler:2007}
{V{\"o}gler}, A., \& {Sch{\"u}ssler}, M. 2007, \aap, 465, L43

\end{thebibliography}

\appendix
\setcounter{figure}{8}
\renewcommand{\thefigure}{\arabic{figure}}

\section{Additional examples of exploding granules}
\label{App:A}
\begin{figure}
  	\centering
   	\resizebox{0.7\hsize}{!}{\includegraphics{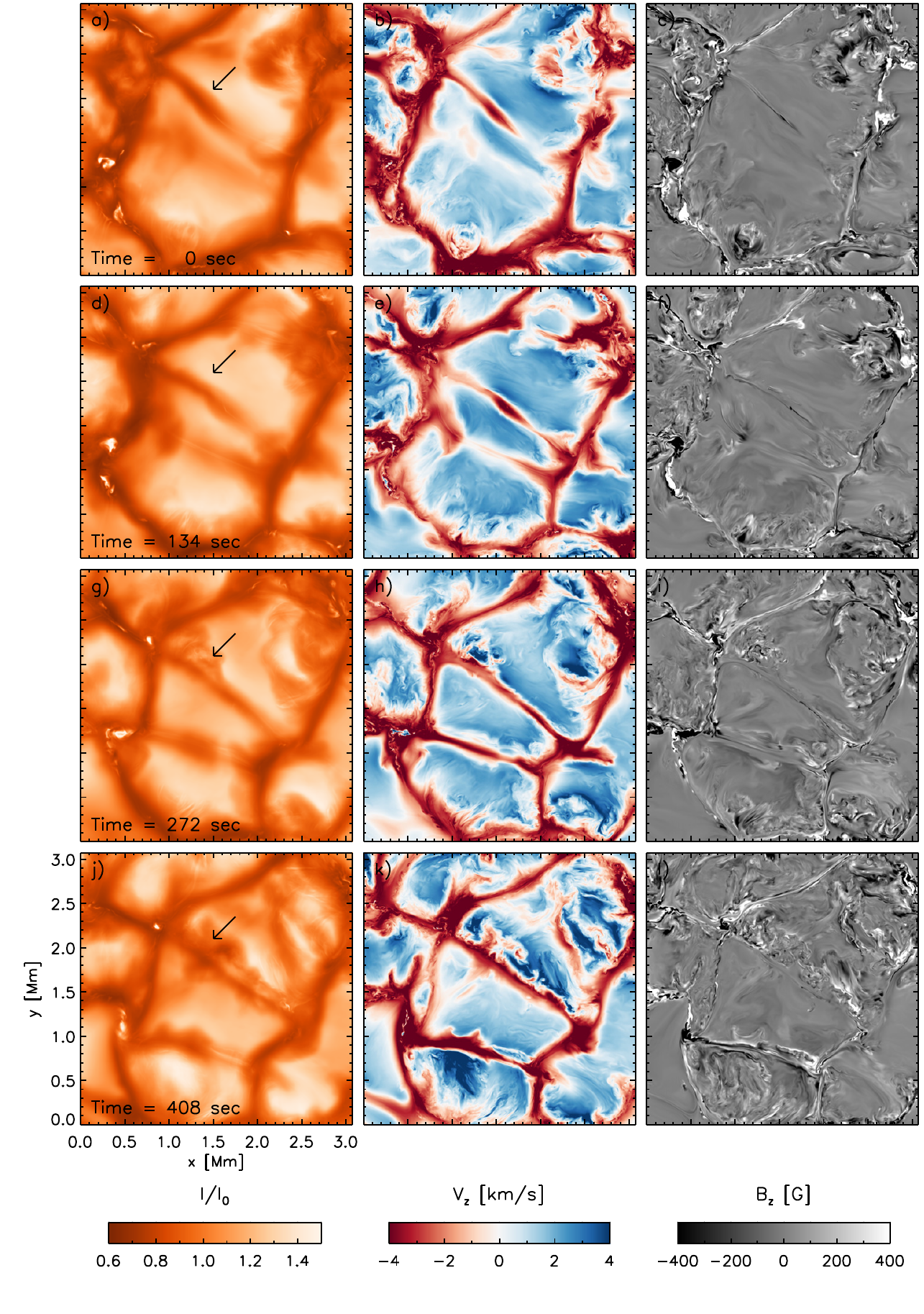}}
   	\caption{Same as Figure \ref{fig:1} for a different case.}
   	\label{fig:9}
\end{figure}

\begin{figure}
  	\centering
   	\resizebox{0.7\hsize}{!}{\includegraphics{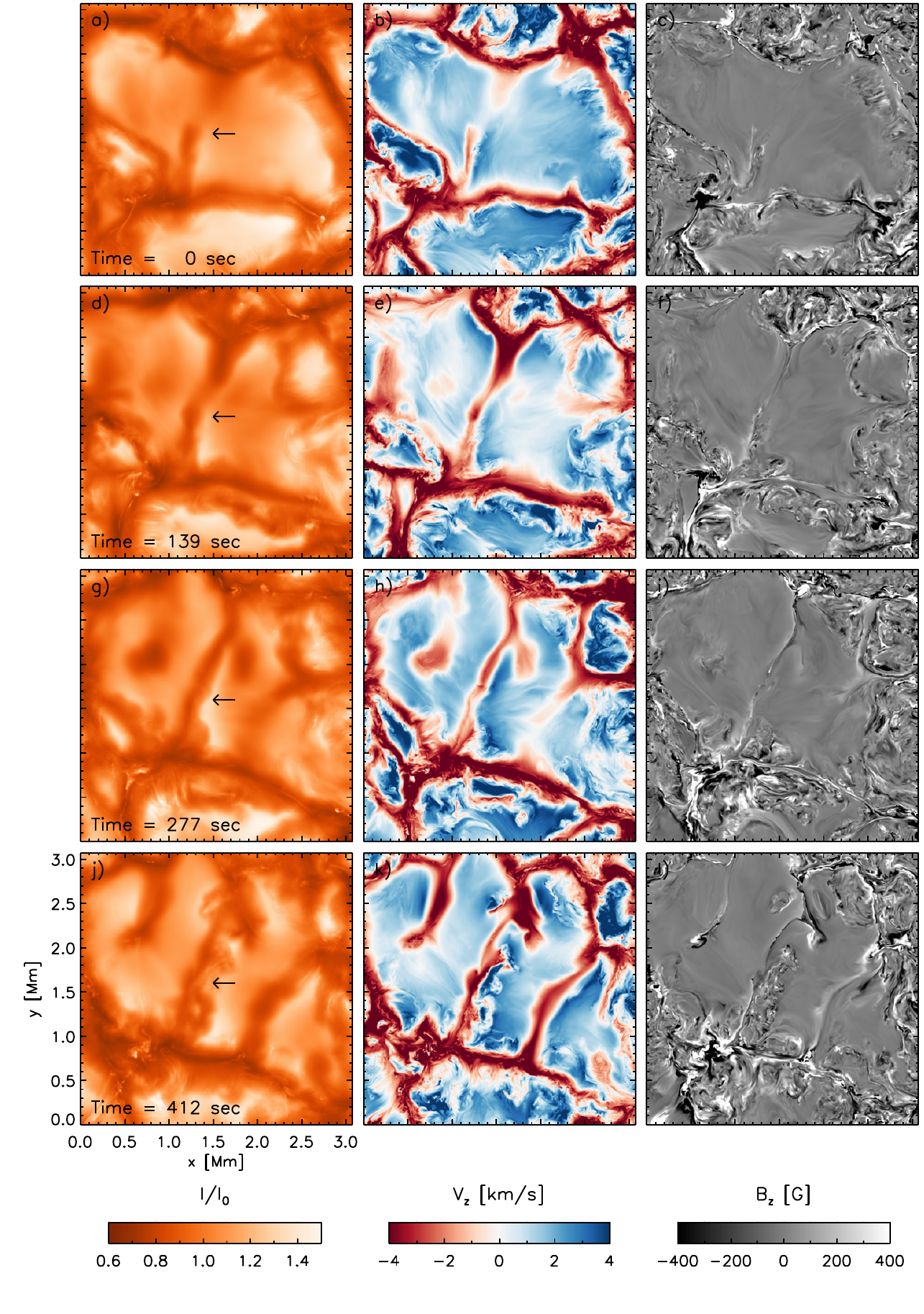}}
   	\caption{Same as Figure \ref{fig:1} for a different case.}
   	\label{fig:10}
\end{figure}

\begin{figure}
  	\centering
   	\resizebox{0.7\hsize}{!}{\includegraphics{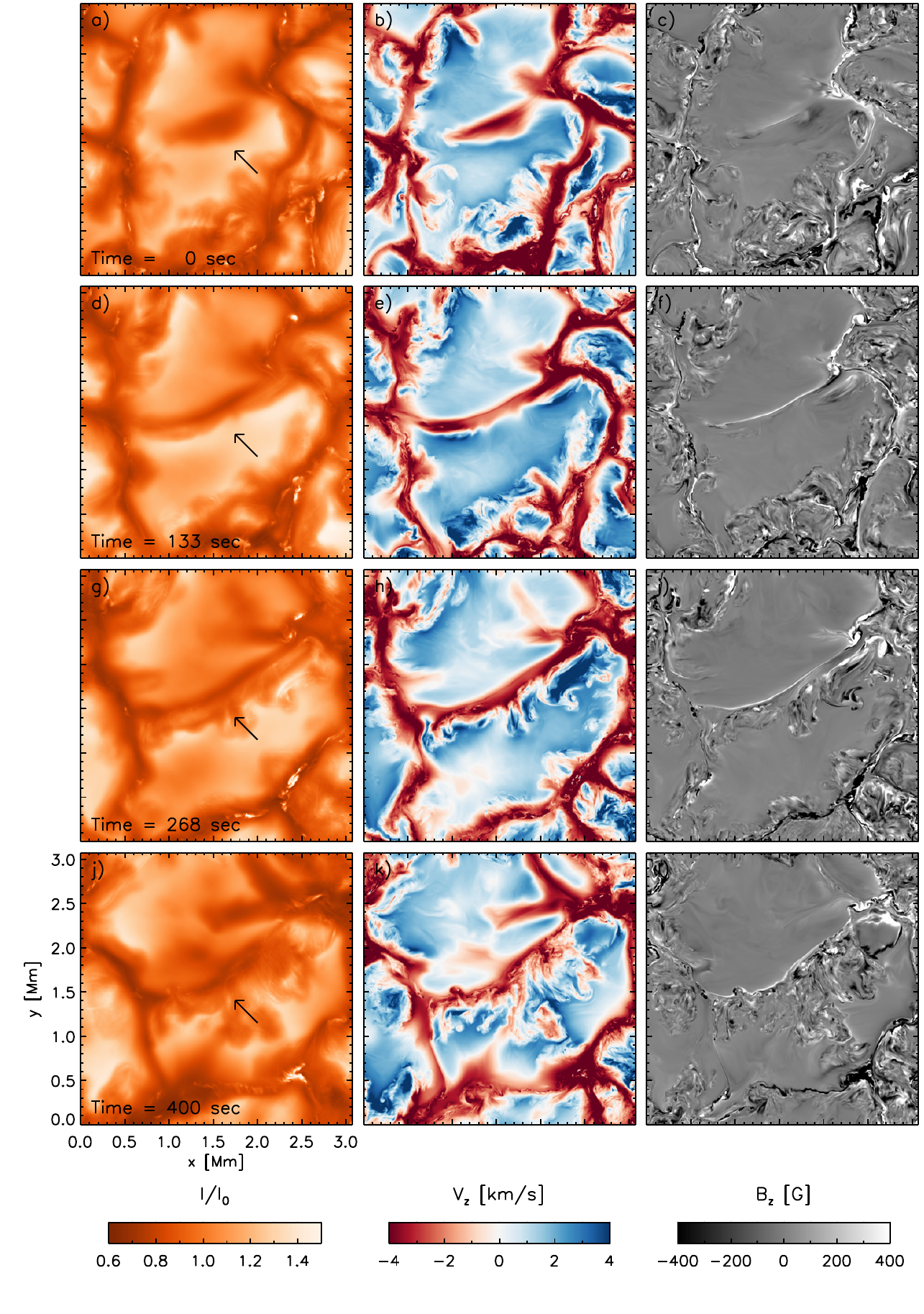}}
   	\caption{Same as Figure \ref{fig:1} for a different case.}
   	\label{fig:11}
\end{figure}

Figures \ref{fig:9}, \ref{fig:10}, and \ref{fig:11} show quantities similar to Figure \ref{fig:1} for three additional examples of exploding granules. These examples
show the same sequence of events: (1) amplification of vertical magnetic field due to convergent horizontal flows leading to concentrated 
magnetic sheets, (2) asymmetric horizontal vorticity leading to a bright edge on one side of the newly formed downflow lane, (3) turbulent
magnetic field appearing in the granular upflow on the edge that showed previously the brighter intensity. 

\NEW{
\section{Numerical diffusivities}
\label{App:B}
We provide a description of numerical diffusivities and computation of viscous and resistive heating as given in
\citet{Rempel:2014:SSD,Rempel:2017:corona}. 
For simplicity we consider here only a one-dimensional problem. The computation of numerical diffusivities is performed in a dimensional
split fashion, i.e. contributions from the 3 grid directions are added sequentially.
The first step is a piece linear reconstruction of the solution $u$ leading extrapolated values at cell interfaces
given by
\begin{eqnarray}
u_l&=&u_i+0.5\,\Delta u_i\\
u_r&=&u_{i+1}-0.5\,\Delta u_{i+1}\;.
\end{eqnarray}
The scheme is applied to the variables $u=\{\log(\varrho), v_x, v_y, v_z, \log{\varepsilon}, B_x, B_y, B_z\}$, where $\varepsilon$ is the 
internal energy per unit mass. Here $u_l$ ($u_r$) are the extrapolated values from cells on the left (right), the reconstruction slope $\Delta u_i$ is 
computed using the monotonized central difference limiter:
\begin{eqnarray}
  \Delta u_i&=&\mbox{minmod}\left[(u_{i+1}-u_{i-1})/2,2\,(u_{i+1}-u_i),2\,(u_i-u_{i-1})\right]\;.
\end{eqnarray}
Numerical diffusive fluxes at cell interfaces are computed from the extrapolated values through the expression
\begin{equation}
  f^{\rm NUM}_{i+\frac{1}{2}}=-\frac{1}{2}\,c_{i+\frac{1}{2}}\,
  \Phi_h(u_r-u_l,u_{i+1}-u_i)\cdot(u_r-u_l)\;,
\end{equation}
where $c_{i+\frac{1}{2}}$ is the characteristic velocity at the cell interface. In the simulations presented here we use
$c=\vert\vec{v}\vert+\sqrt{(0.2\, C_S)^2+V_A^2}$. The function $\Phi_h$ allows to further control the (hyper-) diffusive character of the scheme, it is given
by
\begin{equation}
  \Phi_h=\mbox{max}\left[0,1+h\left(\frac{u_r-u_l}{u_{i+1}-u_i}-1\right)\right]
  \label{phi_lim}
\end{equation}
in regions with $(u_r-u_l)\cdot(u_{i+1}-u_{i})>0$, while $\Phi_h=0$ if $(u_r-u_l)\cdot(u_{i+1}-u_{i})\leq 0$ (no anti-diffusion). For $h=0$ the scheme reduces the diffusive 
flux to that of a standard second order Lax-Friedrichs scheme, while for $h>0$ the diffusivity is reduced for smooth regions in which $\vert(u_r-u_l)/(u_{i+1}-u_{i})\vert<1$.
Some additional contributions from a $4^{th}$ order hyper-diffusion as well as correction terms for mass diffusion as detailed in \citet{Rempel:2014:SSD} are added,
but their contributions are insignificant and not explicitly spelled out here.
}

\NEW{
Applying the scheme to the velocity $v_i^m$ and magnetic field $B_i^m$ variables ($i$ denotes the grid position, whereas $m=1,2,3$ the vector components) leads to the following 
expressions for numerical viscous and resistive heating:
}

\NEW{
\begin{eqnarray}
        \left(Q^{\rm NUM}_{visc}\right)_i&=&\frac{1}{2}\varrho_i\sum_{m=1}^3\left[\left(f^{\rm NUM}_v\right)_{i-{1\over 2}}^m\frac{v_i^m-v_{i-1}^m}{\Delta x}+\left(f^{\rm NUM}_v\right)_{i+{1\over 2}}^m\frac{v_{i+1}^m-v_{i}^m}{\Delta x}\right]\label{Eq:Qvis}\\
         \left(Q^{\rm NUM}_{res}\right)_i&=&\frac{1}{8\pi}\sum_{m=1}^3\left[\left(f^{\rm NUM}_B\right)_{i-{1\over 2}}^m\frac{B_i^m-B_{i-1}^m}{\Delta x}+\left(f^{\rm NUM}_B\right)_{i+{1\over 2}}^m\frac{B_{i+1}^m-B_{i}^m}{\Delta x}\right]\label{Eq:Qres}\;.
\end{eqnarray}
}

\end{document}